\documentclass[10pt,a4paper]{book}
\usepackage{udineHyderabad-Klu2}

\begin{document}

	\pagestyle{fancy}

\talktitle{The MAGIC experiment and its first results}{The MAGIC experiment and its first results\vspace*{2mm}}

\talkauthors{D.~Bastieri\structure{a}, R.~Bavikadi\structure{b}, C.~Bigongiari\structure{a}, E.~Bisesi\structure{b}, P.~Boinee\structure{b},
A.~De~Angelis\structure{b}, B.~De~Lotto\structure{b}, A.~Forti\structure{b}, T.~Lenisa\structure{b}, F.~Longo\structure{c}, O.~Mansutti\structure{b},
M.~Mariotti\structure{a}, A.~Moralejo\structure{a}, D.~Pascoli\structure{a}, L.~Peruzzo\structure{a}, A.~Saggion\structure{a}, P.~Sartori\structure{a},
V.~Scalzotto\structure{a} and\\ 
The MAGIC collaboration}

\authorstucture[a]{Dipartimento di Fisica dell'Universi\`a di Padova and INFN, Italy}

\authorstucture[b]{Dipartimento di Fisica dell'Universi\`a di Udine and INFN, Italy}

\authorstucture[c]{Dipartimento di Fisica dell'Universi\`a di Trieste and INFN, Italy}

\shorttitle{Status of MAGIC}

\firstauthor{MAGIC Collaboration}

		\index{De Lotto@\textsc{De Lotto}, B.}
		\index{Bastieri@\textsc{Bastieri}, D.}
		\index{Bavikadi@\textsc{Bavikadi}, S.R.}
		\index{Bigongiari@\textsc{Bigongiari}, C.}
		\index{Bisesi@\textsc{Bisesi}, E.}
		\index{Boinee@\textsc{Boinee}, P.}
		\index{De Angelis@\textsc{De Angelis}, A.}
		\index{Forti@\textsc{Forti}, A.}
		\index{Lenisa@\textsc{Lenisa}, T.}
		\index{Longo@\textsc{Longo}, F.}
		\index{Mansutti@\textsc{Mansutti}, O.}
		\index{Mariotti@\textsc{Mariotti}, M.}
		\index{Moralejo@\textsc{Moralejo}, A.}
		\index{Pascoli@\textsc{Pascoli}, D.}
		\index{Peruzzo@\textsc{Peruzzo}, L.}
		\index{Saggion@\textsc{Saggion}, A.}
		\index{Sartori@\textsc{Sartori}, P.}
		\index{Scalzotto@\textsc{Scalzotto}, V.}
		\index{MAGIC Collaboration@\textsc{MAGIC Collaboration}}

\begin{abstract}
With its diameter of 17m, the MAGIC telescope is the largest Cherenkov detector for gamma ray astrophysics. It is sensitive to photons above an energy of 30 GeV.
MAGIC started operations in October 2003 and is currently taking data. This report summarizes its main characteristics, its first results and its potential for physics.
\end{abstract}

\section{Introduction}
The MAGIC (Major Atmospheric Gamma Imaging Cherenkov) telescope was designed in 1998 \protect\cite{Barrio} with the main goal of being the Imaging Atmospheric Cherenkov Telescope (IACT) with the lowest gamma energy threshold. It is based on the experience acquired with the first generation of Cherenkov telescopes, and includes a large number of technological improvements\protect\cite{Mirzoyan}.
With a  reflector diameter of 17~m, it is the largest Imaging Cherenkov Telescope in the world.

The study of $\gamma$ rays is fundamental for our understanding of the universe\protect\cite{dea-reviews}: $\gamma$ rays probe the most energetic phenomena occurring in nature, and several signatures of new physics are associated with the emission of $\gamma$ rays. 
Photons can travel essentially undeflected and unabsorbed in space, and thus they point with excellent approximation to the source of their emission.

Together with HESS \protect\cite{hess} in Namibia, VERITAS \protect\cite{veritas} in the USA, and CANGAROO \protect\cite{cangaroo} in Australia, MAGIC is one of the ``big four'' ground-based gamma experiments; among them, it is the only one consisting for the moment of a single telescope.

MAGIC comes after the scientific success of the Energetic Gamma Ray Experiment Telescope \hspace*{-0.8pt}(\hspace*{-0.7pt}EGRET\hspace*{-0.7pt})\hspace*{-0.8pt} instrument on the Compton Gamma Ray Observatory\protect\cite{EGRET}. Launched in 1991, EGRET made the first complete survey of the sky above
30 MeV. EGRET increased the number of identified $\gamma$ sources producing a catalog which is a reference. Still a large fraction of EGRET sources are unidentified and, besides other fundamental physics goals, MAGIC aims at their identification.

MAGIC is located at the Roque de los Muchachos Observatory (ORM) at 2200 m asl (28.8$^o$ north,
17.9$^o$ west) on the Canary island of La Palma.


MAGIC has an effective area of about 4$\cdot 10^4$ m$^2$,  angular resolution of about 0.2 degrees, relative energy resolution of the order of 20\% and can well separate gammas from  background (mainly due to hadrons).

The MAGIC detector is shown in Figure~1.
\begin{figure}
\begin{center}
\includegraphics[width=0.5\columnwidth]{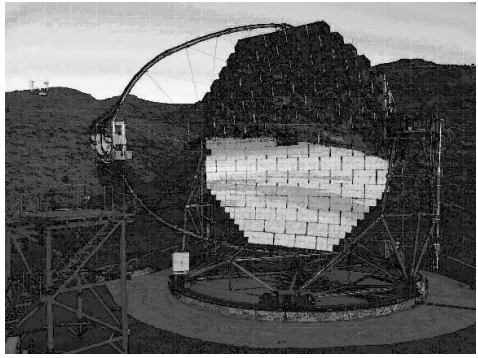}\end{center}
  \caption{The MAGIC\label{foto} detector.}
\end{figure}

The sensitivity of MAGIC as calculated from Monte Carlo is shown together with the expected sensitivity from other gamma-ray detectors in the GeV-TeV range in Figure~2.
\begin{figure}
\begin{center}
\includegraphics[width=0.5\columnwidth]{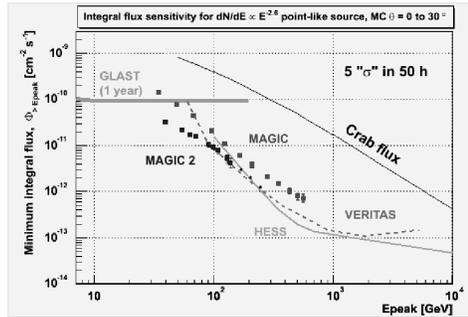}\end{center}
  \caption{Sensitivities\label{sensi} for some operating and proposed gamma detectors.}
\end{figure}

MAGIC has already observed a source, 1ES1959, 10 times fainter than Crab.

By observing gamma rays in the energy range from a few tenth of GeV upwards, in  overlap with satellite observations and with a substantial improvement in sensitivity, energy and angular resolution, results in several fields of science can be achieved:
\begin{itemize}
\item Fundamental physics.
\begin{itemize}
 \item Gamma Ray Horizon, with a larger sensitivity for tests of light propagation effects.
 \item  Dark matter, extending the sensitivity to many theoretical models.
 \item  Quantum gravity: the search for effects using time delays in the arrival of gammas dependent on the energy, will profit of an improved energy resolution.
\end{itemize}
\item Astrophysics.
\begin{itemize}
 \item  Gamma Ray Bursts (GRBs): several tests of fundamental physics can be performed based on the  characteristics of the transient emission. GRBs are among the most distant and powerful sources in the Universe and key informations on space-time structure and cosmology can be gained. In addition a more accurate localization of such sources can help in correlating GRBs with known astrophysical objects.
 \item  Supernova remnants and plerions: improved energy resolution will help in discriminating
between the various acceleration mechanisms assumed to be at the origin of VHE gammas.
 \item  Active Galactic Nuclei: the precise gamma-ray observation of more AGNs at different redshifts will much contribute to answer one of the major open questions in extragalactic astronomy, the formation and evolution of galaxies in the early universe.
 \item  Pulsars: the known pulsars have cutoff energies of their pulsed emission in the few GeV range, hence their detection will become possible by lowering the energy threshold. 
 \item  Unidentified EGRET sources: this is an enormously rich field of activity for detailed studies, possible with modest observation times on nearly half of the observable EGRET sources.
 \item  Diffuse photon background: the present knowledge of both the extragalactic background radiation and the diffuse galactic emission would improve with a more accurate pointing, which could help in separating unidentified sources from a continuum emission.
 \item  Nearby galaxies: their expected steep energy spectrum makes observations at low energy particularly important, as they allow enough flux to be detected in the gamma ray domain.
\item Studies on the galactic center, where a precise pointing might help in reducing the background on nearby sources.
\end{itemize}
\end{itemize}

\section{First results}
 
All the results presented here are preliminary since the analysis software is still in its development phase.

In the first months of data taking, we have mainly concentrated on low zenith angle ($<$40$^{\circ}$) 
observations of TeV sources like Crab Nebula, Mrk~421, Mrk~501, 1ES~1426 and 1ES~1959.
A roughly equivalent amount of OFF source data have been recorded under the same conditions as the ON data for background substraction.

We have applied standard analysis based on the angle $\alpha$ between the major axis of the ellipsis which gives the best approximation of the energy deposition in the detector and the line joining the tenter of such ellipsis to the center of the camera\protect\cite{hillas}.

We give in what follows a flavor of the results collected during the first semester of 1004 on two sources: Crab Nebula and Markarian 421.

\subsection{Crab Nebula}

The Crab nebula is one of the most studied sources in the sky. It radiates a wide range of electromagnetic radiation from radio to 100 TeV $\gamma$ rays.  The spectrum of this source has been measured in the GeV range by EGRET and at energies above 300~GeV by a number of Cherenkov telescopes.  MAGIC can determine for the first time the peak of the emission by inverse Compton, where no experimental data have been published, yet.

A significance of around $\sim$20$\sigma$ per hour of data taking is presently reached by MAGIC.
The data can be subdivided in ranges of energy to obtain an energy scan in the region of interest. Presently we are sensitive to energies above $\simeq 50$ GeV and below 2 TeV (Figure 3).
\begin{figure}
\begin{center}
\includegraphics[width=0.8\columnwidth]{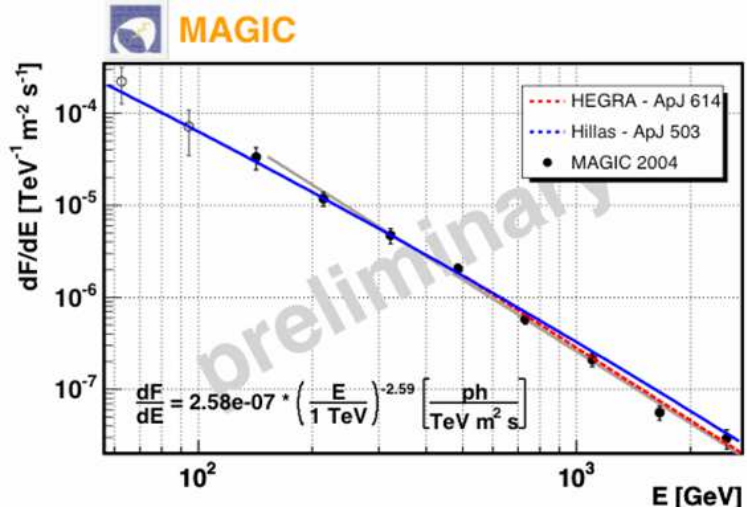}\end{center}
  \caption{Differential energy spectrum\label{crab}  from Crab; data are preliminary. The study of the efficiency on the first two bins is still under study.}
\end{figure}

\subsection{Mrk~421}

Mrk~421 undergoes the fastest flares that have been observed at the TeV energies, with durations as short as 20~minutes. During 2004 the source has undergone an episode of intense flaring.

The preliminary MAGIC data show a significant signal also at energies as low as 50-60 GeV.

\section{Conclusions}
MAGIC's  preliminary results confirm its instrumental capabilities  to scan gamma emissions from the Universe above 30 GeV. MAGIC has demonstrated the observability of sources as faint as 10\% of Crab.

The construction of a second detector (MAGIC II) to be put at 80 m from MAGIC has started, and completion is expected in 2006.

\end{document}